\documentclass[utf8]{FrontiersinHarvard} 

\usepackage{url,hyperref,lineno,microtype,subcaption}
\usepackage{graphicx}
\usepackage{colortbl}
\usepackage{caption}
\usepackage{capt-of}
\usepackage{balance}
\usepackage{physics}
\usepackage{fixltx2e}
\usepackage{amsmath}
\usepackage{array}
\usepackage{multirow}
\usepackage{comment}
\usepackage{enumerate}
\usepackage[onehalfspacing]{setspace}

\def\keyFont{\fontsize{8}{11}\helveticabold }
\def\firstAuthorLast{Giri {et~al.}} 
\def\Authors{Amita Giri\,$^{1,*}$, John C. Mosher\,$^{2}$, Amir Adler\,$^{3,1}$ and Dimitrios Pantazis\,$^{1}$}

\def\Address{$^{1}$McGovern Institute for Brain Research, Massachusetts Institute of Technology, Cambridge, MA, USA \\
$^{2}$Texas Institute for Restorative Neurotechnologies, McGovern Medical School, Department of Neurology, UTHealth, Houston, TX, USA \\
$^{3}$Electrical Engineering Department, Braude College of Engineering, Karmiel, Israel}
\def\corrAuthor{Corresponding Author}

\def\corrEmail{amita3gb@mit.edu}

\begin{document}
\onecolumn
\firstpage{1}

\title {An F-ratio-Based Method for Estimating the Number of Active Sources in MEG} 

\author[\firstAuthorLast ]{\Authors} 
\address{} 
\correspondance{} 

\extraAuth{}

\maketitle

\begin{abstract}

\section{}
Magnetoencephalography (MEG) is a powerful technique for studying the human brain function. However, accurately estimating the number of sources that contribute to the MEG recordings remains a challenging problem due to the low signal-to-noise ratio (SNR), the presence of correlated sources, inaccuracies in head modeling, and variations in individual anatomy. To address these issues, our study introduces a robust method for accurately estimating the number of active sources in the brain based on the F-ratio statistical approach, which allows for a comparison between a full model with a higher number of sources and a reduced model with fewer sources. Using this approach, we developed a formal statistical procedure that sequentially increases the number of sources in the multiple dipole localization problem until all sources are found. Our results revealed that the selection of thresholds plays a critical role in determining the method’s overall performance, and appropriate thresholds needed to be adjusted for the number of sources and SNR levels, while they remained largely invariant to different inter-source correlations, modeling inaccuracies, and different cortical anatomies. By identifying optimal thresholds and validating our F-ratio-based method in simulated, real phantom, and human MEG data, we demonstrated the superiority of our F-ratio-based method over existing state-of-the-art statistical approaches, such as the Akaike Information Criterion (AIC) and Minimum Description Length (MDL). Overall, when tuned for optimal selection of thresholds, our method offers researchers a precise tool to estimate the true number of active brain sources and accurately model brain function. 

\tiny
 \keyFont{ \section{Keywords:} F-ratio, Source Localization, Alternating Projection (AP), Source Enumeration, MEG, AIC, MDL} 
\end{abstract}

\section{Introduction}
Magnetoencephalography (MEG) is a powerful non-invasive neuroimaging technique that offers high temporal resolution for studying human brain function \citep{hamalainen1993magnetoencephalography,baillet2017magnetoencephalography}. Localization of MEG sources has garnered significant interest in recent years since it can reveal the origins of neural signals and offer valuable insights into the complex workings of the human brain. By identifying the sources of neural activity, researchers can study the underlying mechanisms of cognition, perception, and other brain functions \citep{ahveninen2006task,giorgetta2013waves, klepp2015language, pancholi2022source}. Additionally, source localization can aid in diagnosing and studying neurological disorders and identifying abnormal brain activity \cite{oishi2002epileptic, PA_9314203,wilkinson2020predicting, westlake2012resting,xu2021graph, giri2022spatial, giri2022anatomical}.

MEG source localization methods typically involve solving an inverse problem, which entails estimating current sources within the brain based on the measured MEG data. This is challenging because the measured signals are influenced by several factors, such as the geometry and conductivity of the head, sensor noise, and the ill-posed nature of the problem. Mathematically, the localization problem can be cast as finding the location and moment of the set of dipoles whose field best matches the M/EEG measurements \cite{mosher_multiple_1992}. Localization methods can be broadly categorized into \textit{distributed} and \textit{discrete} solutions. 

\textit{Distributed source imaging} approaches aim to estimate a density map of active dipoles across the entire cortex. Commonly used methods include MNE \cite{hamalainen_magnetoencephalographytheory_1993, hamalainen1994interpreting}, dSPM \cite{MinNormNN}, and sLORETA \cite{pascual2002standardized}. However, these methods assume a significantly larger number of unknown sources in a discrete surface or volumetric grid compared to the number of MEG sensors. The ill-posed nature of the problem poses a significant challenge, especially in the presence of multiple active regions in the brain \cite{darvas2004mapping}. Non-linear source estimation methods, such as MxNE \cite{strohmeier2016iterative} and TF-MxNE \cite{gramfort2013time}, address this issue by incorporating $l1$-norm penalty regularizers that favor sparse collections of focal dipolar sources. While these methods have shown some success, they tend to be computationally demanding and have limited accuracy when dealing with complex multi-dipole configurations. 

On the other hand, \textit{discrete multiple dipole localization} methods avoid the ill-posedness associated with distributed methods by finding a small set of equivalent current dipoles (ECDs) whose field best matches the M/EEG measurements in a least-squares sense \cite{mosher_multiple_1992}. Dipole localization methods offer a more classical approach to brain source localization and provide more intuitive interpretations of brain activity by estimating the location, orientation and amplitude of neural sources. The most well-known methods are beamformers \cite{van1997localization, vrba2001signal} and MUSIC \cite{mosher_multiple_1992}, and their recursive variants RAP-MUSIC \cite{Mosher1999}, Truncated RAP-MUSIC \cite{makela_truncated_2018}, and RAP Beamformer \cite{ilmoniemi2019brain}. While recursive variants generally perform better than their non-recursive counterparts, they still face limitations such as reduced effectiveness, reliance on high signal-to-noise ratio (SNR), and potential cancellation of correlated sources. Recent advancements in this field have addressed some of these concerns, including Alternating Projections (AP) \citep{adler2022brain}, DS-MUSIC \citep{makela2017locating,ilmoniemi2019brain}, and Flex-MUSIC \citep{hecker2023source}. Estimating the number of independent signal components is a prerequisite for dipole localization methods to accurately estimate dipole sources. However, determining the correct number of active sources contributing to the recorded signals remains a fundamental challenge in MEG data analysis \citep{wendel2009eeg}, significantly impacting the success of brain source localization. We focus on addressing this specific problem here.

Estimating the number of active sources in MEG data poses significant challenges due to multiple factors. First, MEG signals generally exhibit a low signal-to-noise ratio (SNR), which makes it difficult to differentiate between simultaneously active sources. Second, the presence of correlated sources adds complexity by potentially causing multiple sources to be mistakenly identified as a single source. Last, errors in head modeling and variations in individual anatomy introduce additional noise and variability, hampering accurate estimation of the location and strength of underlying sources.

Early attempts to estimate the number of dipoles relied on subjective thresholds \citep{bartlett1954note, lawley1956tests, chen1991detection}. These approaches involved setting a threshold that separated the eigenvalues of the data covariance matrix from the complete set of eigenvalues. Chen \citep{chen1991detection} proposed a method that detected the number of sources by imposing an upper bound on the eigenvalue magnitudes of the correlation matrix derived from the array output. In addition to conventional eigenvalue-based techniques, a few methods have also employed eigenvectors for estimating the number of sources \citep{di1984matrix, jiang2004robust}.

To overcome the limitations of subjective thresholds, two main classes of methods have been developed for estimating the number of signal sources using the distribution of the eigenvalues of the data covariance matrix. The first class involves techniques based on principal component analysis (PCA) \citep{green1988transformation, yao2018estimating}, independent component analysis (ICA) \citep{ikeda2000independent}, and factor analysis \citep{malinowski1977determination, malinowski1977theory}. The second class consists of information theoretic approaches \citep{knosche1998determining, wax1985detection}, such as the Akaike information criterion (AIC) \citep{akaike1974new} and the minimum description length (MDL) \citep{schwarz1978estimating, rissanen1978modeling}. The work from \citep{wax1985detection} derived the eigenvalue forms of AIC and MDL methods, which can be directly applied to array signal processing problems. These methods aim to strike a balance between model fit and complexity using principles from information theory. However, these approaches assume source independence, which is not always valid in the brain. As a result, they tend to perform poorly \citep{chen1991detection, zhang1989statistical, yao2018estimating, salman2015estimating}, especially in the presence of correlated sources, noise, low SNR, and limited time samples. Hence, there is a need for more robust and accurate methods to estimate the number of active sources in MEG signals.

A study by \citep{supek1993simulation} aimed to evaluate the efficacy of three statistical measures, namely percent of variance, reduced chi-square, and F-ratio, in determining the correct number of sources (model order). The authors advocated for the reduced chi-square method as a reliable measure of goodness-of-fit, whereas they were less favorable to the percent of variance and F-ratio because they ignored noise contributions. Although we agree that the percent of variance has limited utility, we contend that the efficacy of the F-ratio method was underestimated. Indeed, the simulation results presented in \citep{supek1993simulation} demonstrated that the F-ratio remained stable across different noise levels and successfully identified the true number of sources. However, it is important to note that the study was confined to simulations on a simple spherical head model, lacked assessments using real data, and did not provide clear threshold decision criteria for determining the correct number of sources.

To address these limitations, we propose a robust method for accurately estimating the number of active sources in the brain using the F-ratio statistical approach. Our method introduces formal decision criteria that sequentially increase the number of sources in the multiple dipole localization problem until all sources are found. Our method is based on the F-ratio test, which is commonly used in statistics to compare the variances of two samples. It is sensitive to differences in the variances of the samples and can be used to determine whether adding a source to the model significantly improves the fit of the model to the data. The F-ratio statistical approach allows for a comparison between a full model with a higher number of sources and a reduced model with fewer sources. 

We validated the F-ratio-based method on simulated, real phantom, and human MEG data, and compared its performance to that of other state-of-the-art statistical approaches, such as AIC and MDL. We found that the F-ratio-based method outperformed competing methods in terms of accuracy and reliability. One crucial aspect we investigated was the selection of appropriate thresholds for the F-ratio values. We found that this selection played a critical role in determining the overall performance of our method. Through systematic analyses, we identified optimal thresholds that needed to be adjusted according to the number of sources and SNR levels. Importantly, these thresholds exhibited remarkable consistency across different inter-source correlations, modeling inaccuracies, and cortical anatomies. When fine-tuned with the optimal selection of thresholds, our F-ratio-based method emerged as a precise and robust tool for estimating the true number of active sources in MEG data. 

\section{Materials and Methods}

In this section, we provide a concise overview of the notations used to describe the measurement data, forward matrix, and sources. We also present the problem formulation for estimating multiple ECDs in the brain. Subsequently, we describe the F-ratio statistical procedure, which serves as the foundation to estimate the number of active sources in the brain, and outline the experimental procedures we use to assess performance.

\subsection{Measurement model and notations}
Consider an array of $M$ MEG sensors detecting signals from $Q$ ECD sources located at positions $\{\mathbf{p}_q\}_{q=1}^{Q}$. At time $t$, the MEG measurement vector $\mathbf{y}(t)$ can be described as a superposition of the contributions from $Q$ source signals $\{s_q(t)\}_{q=1}^{Q}$ and additive noise:

\begin{equation}
 \mathbf{y}(t) = \sum _{q=1}^{Q} \mathbf{l}(\mathbf{p}_q)s_q(t) + \mathbf{n}(t), ~~~\text{where}~~ Q < M
\end{equation}
The topography $\mathbf{l}(\mathbf{p}_q)$ of the $q$th dipole at location $\mathbf{p}_q$ is defined as $ \mathbf{l}(\mathbf{p}_q) = \mathbf{L}(\mathbf{p}_q) \mathbf{o}$, where $\mathbf{L}(\mathbf{p}_q) \in \mathbb{R}^{M \times 3}$ is the lead field matrix and $\mathbf{o}\in \mathbb{R}^{3 \times 1}$ is the orientation vector. Depending on the problem, the orientation vector $\mathbf{o}$ may either be known, referred to as a \textit{fixed-oriented} dipole, or it may be unknown, referred to as \textit{freely-oriented}. Additionally, the measurements are subject to the presence of additive white Gaussian noise, which is represented by $\mathbf{n}(t) \in \mathbb{R}^{M \times 1}$.

Several source localization methods exist for estimating the $Q$ ECD sources, with each dipole source characterized by its location, orientation, and amplitude. Example methods include MUSIC \citep{mosher_multiple_1992}, RAP-MUSIC \cite{Mosher1999}, Truncated RAP-MUSIC \cite{makela_truncated_2018}, RAP Beamformer \cite{ilmoniemi2019brain}, HSH-MUSIC \citep{giri2018eeg}, H$^2$-MUSIC \cite{giri2019head}, DS-MUSIC \citep{makela2017locating,ilmoniemi2019brain}, and Flex-MUSIC \citep{hecker2023source}. More recently, we introduced a method called Alternating Projections (AP) \citep{adler2022brain}, which offers several advantages. AP source localization method is robust to forward model errors, can handle high inter-source correlation values, and is effective even in low signal-to-noise ratio (SNR) scenarios. It is important to note that estimating the true number of active sources $Q$ is a fundamental requirement for all the aforementioned dipole localization methods to accurately estimate the dipole source parameters.

\subsection{F-ratio based method}

The F-test is a widely used statistical technique that leverages the F-ratio to assess the presence of a significant difference between the variances of two data sets. In the context of determining the true number of sources, this technique holds particular value. It enables us to test the hypothesis that incorporating an additional source results in a substantial improvement in the variance accounted for by the model. By employing the F-test, we can make informed decisions regarding the optimal number of sources to include in order to provide the most accurate explanation for the observed data.

In probability theory and statistics, the F-statistic, also known as the F-ratio, is defined as the ratio between two independent chi-square distributions, denoted as $\mathbf{X_1} \sim \mathbf{\chi}^{2}_{\text{DOF}_\text{1}}$ and $\mathbf{X_2} \sim \mathbf{\chi}^{2}_{\text{DOF}_\text{2}}$, where ${\text{DOF}_\text{1}}$ and ${\text{DOF}_\text{2}}$ represent the respective degrees of freedom. Mathematically, the F-ratio is expressed as:
\begin{equation}
\mathbf{F} = \frac{\mathbf{X_1}/{\text{DOF}_\text{1}}}{\mathbf{X_2}/{\text{DOF}_\text{2}}}
\label{eq:08}
\end{equation}
This formula provides a means to calculate the F-ratio by dividing the observed value of $\mathbf{X_1}$ normalized by its degrees of freedom, ${\text{DOF}_\text{1}}$, by the observed value of $\mathbf{X_2}$ normalized by its degrees of freedom, ${\text{DOF}_\text{2}}$. In this study, we use this idea to compare two hypothesized models based on the variance they explain. The first model, referred to as the ``reduced model", explains the data with $K$ number of sources:
\begin{align}
 \mathbf{y}(t) &= \mathbf{y}_{\text{R}}(t) + \mathbf{n}_{\text{R}}(t), \\
 \mathbf{y}_{\text{R}}(t) &= \sum _{q=1}^{K} \mathbf{l}(\mathbf{p}_q){s_{q}}_{\text{R}}(t)
\end{align}
where $\mathbf{y}_{\text{R}}(t)$ represent the estimated signals and $\mathbf{n}_{\text{R}}(t)$ represents noise of a reduced model. On the other hand, the second model, called the ``full model," includes one more source compared to the reduced model, with $K+1$ sources:
\begin{align}
 \mathbf{y}(t) &= \mathbf{y}_{\text{F}}(t) + \mathbf{n}_{\text{F}}(t), \\
 \mathbf{y}_{\text{F}}(t) &= \sum _{q=1}^{K+1} \mathbf{l}(\mathbf{p}_q){s_{q}}_{\text{F}}(t)
\end{align}
where $\mathbf{y}_{\text{F}}(t)$ represent the estimated signals and $\mathbf{n}_{\text{F}}(t)$ represents noise of a full model. The estimation of $\mathbf{y}_{\text{R}}(t)$ and $\mathbf{y}_{\text{F}}(t)$ signals is achieved by solving an inverse problem using a dipole fitting method.
\begin{figure}[t]
\begin{center}
\includegraphics[width=0.95\linewidth]{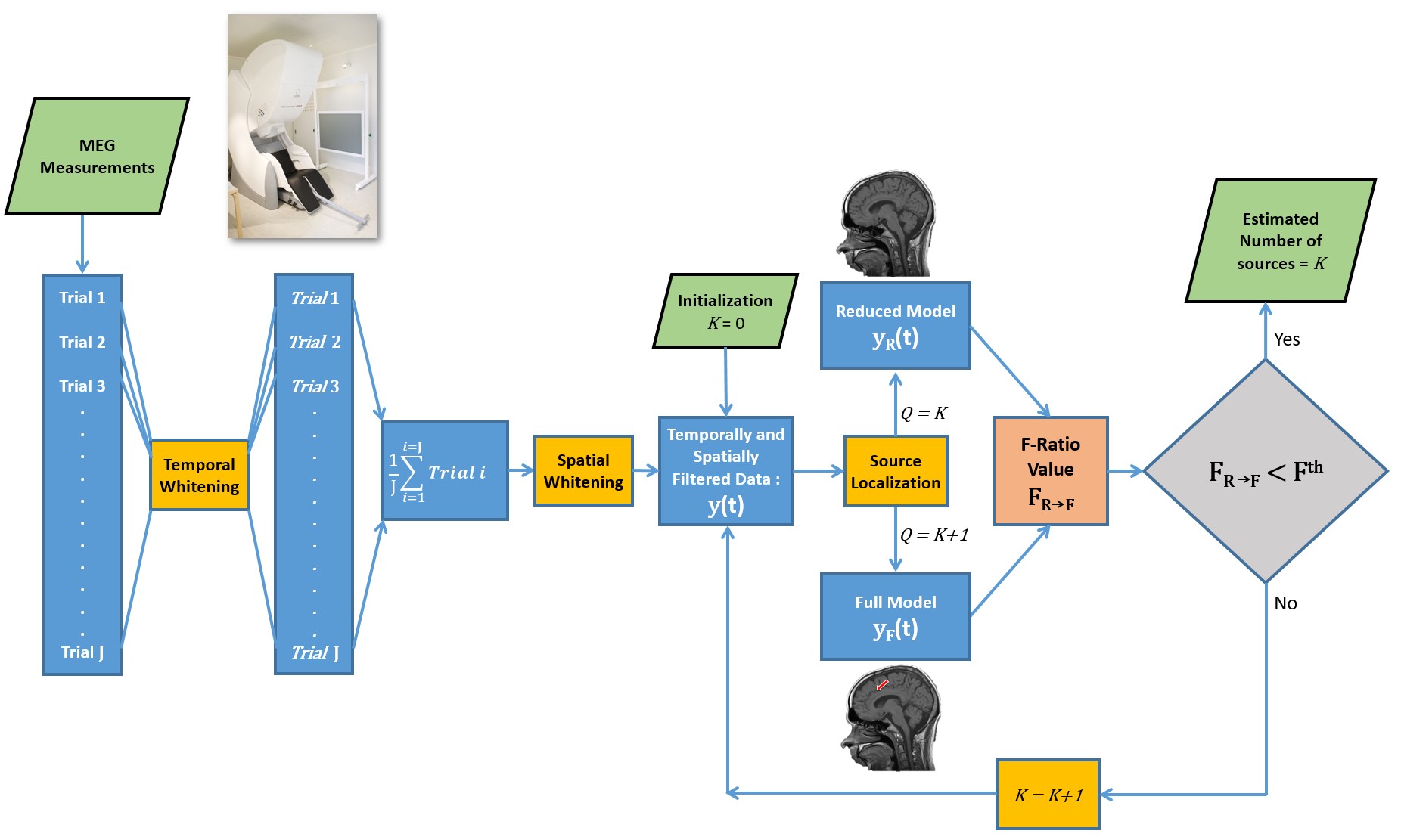}
\end{center}
\caption{Flowchart of the F-ratio method for estimating the number of sources.}
\label{fig:00}
\end{figure}
By comparing the residual variance of these two models, we can assess their performance and determine the most appropriate model for the given data. Since we assume that the noise added to the measured signal is white Gaussian noise, it can be deduced that the sum of square errors between the measured signal $\mathbf{y}(t)$ and the estimated signals $\mathbf{y}_{\text{R}}(t)$ and $\mathbf{y}_{\text{F}}(t)$, denoted as $\norm{\mathbf{y}(t) - \mathbf{y}_{\text{R}(t)}}_2^{2}$ and $\norm{\mathbf{y}(t) - \mathbf{y}_{\text{F}}(t)}_2^{2}$ respectively, follows chi-square distributions. Therefore, the F-ratio test can be written as \citep{supek1993simulation}:
\begin{equation}
 \textbf{F}_{{\text{R} \rightarrow \text{F}}} = \frac{\sum_{t=1}^{N}\norm{\mathbf{y}(t) - \mathbf{y}_{\text{R}}(t)}_2^{2}} {\text{DOF}_\text{R}}~\Bigg{/}~\frac{\sum_{t=1}^{N} \norm{\mathbf{y}(t) - \mathbf{y}_{\text{F}}(t)}_2^{2}}{\text{DOF}_\text{F}}
\end{equation}
where $N$ is the number of time samples. For the fixed-oriented case, the degrees of freedom (DOF) for the reduced and full models are given by $\text{DOF}_\text{R} = MN - (3+N)K$ and $\text{DOF}_\text{F} = MN - (3+N)(K+1)$, respectively. For the freely-oriented case, the DOF for the reduced and full models are $\text{DOF}_\text{R} = MN - (4+N)K$ and $\text{DOF}_\text{F} = MN - (4+N)(K+1)$, respectively. These formulas account for a total of $MN$ degrees of freedom, with three degrees of freedom deducted for position, one for orientation (in the freely-oriented case), and $N$ for amplitude, for each dipole.

Note that the formulas for estimating the DOF assume independence among all data points, which is not the case in experimental data. To address this issue, in experimental data we implemented a two-step whitening process involving both temporal and spatial filtering. The first step involved temporal whitening, as depicted in the flowchart of the F-ratio method (Figure \ref{fig:00}). The MEG data of each trial was whitened temporally using a six-order linear predictive coding (LPC) technique. This method helped alleviate temporal correlations within the data, reducing their impact on the results. Subsequently, the LPC-filtered trials were averaged. The second step was spatial whitening, which was achieved by applying a whitening filter derived from inverting the noise covariance matrix. This step further mitigated inter-dependencies among the data points, enhancing the reliability of the analysis. A more detailed discussion of the two-step whitening process is presented in Section \ref{sec:2.4}.

It is important to note that the calculated F-ratio values are influenced by the residuals obtained after dipole fitting, and therefore, are dependent on the chosen source localization method used for solving the ECD localization problem in MEG. In our study, we specifically examined the behavior of the F-ratio statistical procedure when employing the AP localization method. The AP method solves the inverse problem iteratively and sequentially by minimizing the least-squares (LS) criterion. For a more detailed and comprehensive discussion of the AP method, we refer readers to \citep{adler2022brain}.

To estimate the number of sources, a systematic approach involving the comparison of a reduced model and a full model is illustrated in Figure \ref{fig:00}. The formal comparison begins by initializing the reduced model with zero sources and the full model with one source to compute the F-ratio value. The decision regarding the number of sources involves comparing the resulting F-ratio value, denoted as $\textbf{F}_{{0 \rightarrow 1}}$, with a threshold value, $\textbf{F}^{\textbf{th}}$. If the reduced model is rejected ($\textbf{F}_{{0 \rightarrow 1}} > \textbf{F}^{\textbf{th}}$), indicating evidence of at least one source, the analysis proceeds to compare a reduced model with one source against a full model with two sources, represented as $\textbf{F}_{{1 \rightarrow 2}}$. This sequential process continues, increasing the number of sources, until reaching a step where the reduced model cannot be rejected, providing an estimation of the true number of sources. 

\subsection{Performance evaluation with simulations}

We evaluated the performance of the F-ratio method in diverse simulated scenarios, considering variations in the number of sources, signal-to-noise ratio (SNR) levels, inter-source correlations, modeling errors, and cortical anatomies. 

The SNR was defined as the ratio between the Frobenius norm of the signal-magnetic-field spatiotemporal matrix and that of the noise matrix, following the approach described in \citep{sekihara2001reconstructing}. To quantify inter-source correlation, we employed the Pearson's correlation coefficient. To establish the desired correlation among the sources, we utilized the Cholesky decomposition method. Initially, we generated fundamental cosine signals for each simulated source, with randomized phase and frequencies ranging from 10Hz to 30Hz. Next, we applied the Cholesky decomposition to factorize the symmetric positive definite target correlation matrix into the product of a lower triangular matrix and its conjugate transpose. By multiplying lower triangular matrix with the fundamental cosine signals, we generated a set of correlated dipole waveforms. This procedure ensured that the resulting source time courses closely matched the correlation coefficients specified by the target correlation matrix, thus incorporating the desired inter-source correlations in our simulations.

The sensor array geometry was based on the Megin Triux MEG system, which consists of a 306-channel probe unit with 204 planar gradiometer sensors and 102 magnetometer sensors. The MEG source space geometries were modeled using the cortical manifold extracted from MR data of adult humans, employing Freesurfer \citep{fischl2004sequence}. In our analysis, we used cortical anatomies from four different adult humans. Simulated sources were restricted to approximately 15,000 grid points distributed over the cortex. The reconstructed sources were estimated on a distinct grid of 50,000 points covering the cortex. To avoid the ``inverse crime" problem, where identical parameters are used for data synthesis and inversion in an inverse problem, the simulation and reconstruction grids were non-overlapping with an average distance of 0.7 mm between neighboring points \citep{colton1998inverse}. The forward matrix for both grids was computed using the boundary element method implemented in OpenMEEG \citep{gramfort2010openmeeg} within the BrainStorm software \citep{tadel2011brainstorm}. Simulated MEG data was generated by randomly selecting sources from the simulation grids. Gaussian white noise was then added to the MEG sensors to model instrumentation noise at specified signal-to-noise ratio (SNR) levels. In order to evaluate the effect of head model errors, we introduced translations to the reconstruction grid before computing the forward matrix. Last, we employed the AP method to solve the inverse problem in the \textit{fixed oriented} case. All experiments were conducted with 100 Monte Carlo simulations to ensure statistical robustness.

\subsection{Performance evaluation with a real phantom} \label{sec:2.4}

We assessed the performance of the F-ratio method using the \textit{freely-oriented} dipoles model with phantom data provided in the phantom tutorial \citep{phantomtutorial} of the Brainstorm software \citep{tadel2011brainstorm}. The phantom experiment was conducted using the Megin Neuromag system, which consists of a 306-channel probe unit with 204 planar gradiometer sensors and 102 magnetometer sensors.

The data comprised MEG recordings obtained from the sequential activation of 32 artificial dipoles. To activate the phantom dipoles, an internal signal generator was used along with an external multiplexer box that connected the signal to each individual dipole. Each dipole was activated 20 times with an amplitude of 200 nAm, resulting in a total of 20 trials for each experimental condition. It is important to note that the chosen amplitude of 200 nAm falls within the range typically observed in inter-ictal spikes associated with epilepsy, as observed in raw data \citep{oishi2002epileptic}.

For each dipole and each trial, the MEG data was whitened temporally using a six-order LPC technique. In particular, baseline data of a 200 ms pre-stimulus interval was used to compute the LPC coefficients of sixth order. These coefficients were then averaged across sensors and subsequently applied to the post-stimulus data modeled as an moving average (MA) filter. The purpose of this step was to eliminate temporal dependencies in the post-stimulus data, as observed in the auto-regressive (AR) model of baseline data. Following this, the LPC-filtered post-stimulus measurements were averaged across the 20 trials. In addition to temporal prewhitening, spatial prewhitening was also performed on the average data using a regularized noise covariance matrix in the Brainstorm software \cite{tadel2011brainstorm}. The regularization process included adding an identity matrix scaled to 10\% of the largest eigenvalue of the noise covariance matrix.

To simulate the concurrent activation of multiple sources, we combined averaged data from different dipoles since only one dipole could be activated at a time. To introduce variability and avoid perfect coherence, we added a random delay ranging from 0 to 50 ms for each dipole.

The reconstruction source space was defined as a sphere centered within the MEG sensor array, with a radius of 64.5 mm. It was sampled using a regular volumetric grid of points with a resolution of 2.5 mm, resulting in a total of 56,762 grid points. The forward matrix was estimated based on a single sphere head model using the BrainStorm software \citep{tadel2011brainstorm}. The performance of the F-ratio method was evaluated using the AP method for localizing dipoles in the \textit{freely-oriented} case. 

\subsection{Performance evaluation with human MEG data}

The effectiveness of the F-ratio method in practical scenarios was assessed using human MEG data recorded from a single human participant during an auditory task. Prior to participation, the subject provided written informed consent, and the study was approved by the local ethics committee (Institutional Review Board of the Massachusetts Institute of Technology), following the principles of the Declaration of Helsinki. During the experiment, binaural sounds (beeps) were delivered to the subject using tubal-insert earphones. These auditory stimuli are known to elicit specific brain responses that are localized in the bilateral primary auditory cortex. A total of 166 trials were recorded, with an interstimulus interval of 2150 ms between each auditory stimulus. The MEG data were acquired using a MEGIN Triux MEG system, which includes a 306-channel probe unit consisting of 102 magnetometers and 204 planar gradiometers.

The forward matrix was estimated using BrainStorm based on an overlapping spheres head model. The reconstruction source space volume was sampled adaptively, with a higher density of points near the surface where we expect a higher spatial resolution due to the proximity to the sensors. The density decreased gradually towards the center of the brain, resulting in a total of 33,073 grid points. This adaptive sampling approach allowed us to optimize the spatial resolution based on the specific requirements of the experiment. 

The performance of the F-ratio method was evaluated using the AP method for localizing dipoles in the \textit{freely-oriented} case. Before localization, the raw data underwent prewhitening in both the temporal and spatial domains following the same procedure as in the real phantom data.

\section{Results and Discussion}
\subsection{Optimal modeling of MEG data requires the adjustment of F-ratio nominal thresholds}

To investigate the behavior of the F-ratio method under different experimental conditions, we conducted a thorough simulation analysis. Our objective was to assess the accuracy of estimating the true number of sources by varying the threshold values across various experimental scenarios. These scenarios included different numbers of active sources, varying SNR levels, inter-source correlation values, modeling errors, and cortical anatomies.

\begin{figure}[t]
\begin{center}
\includegraphics[width=0.92\linewidth]{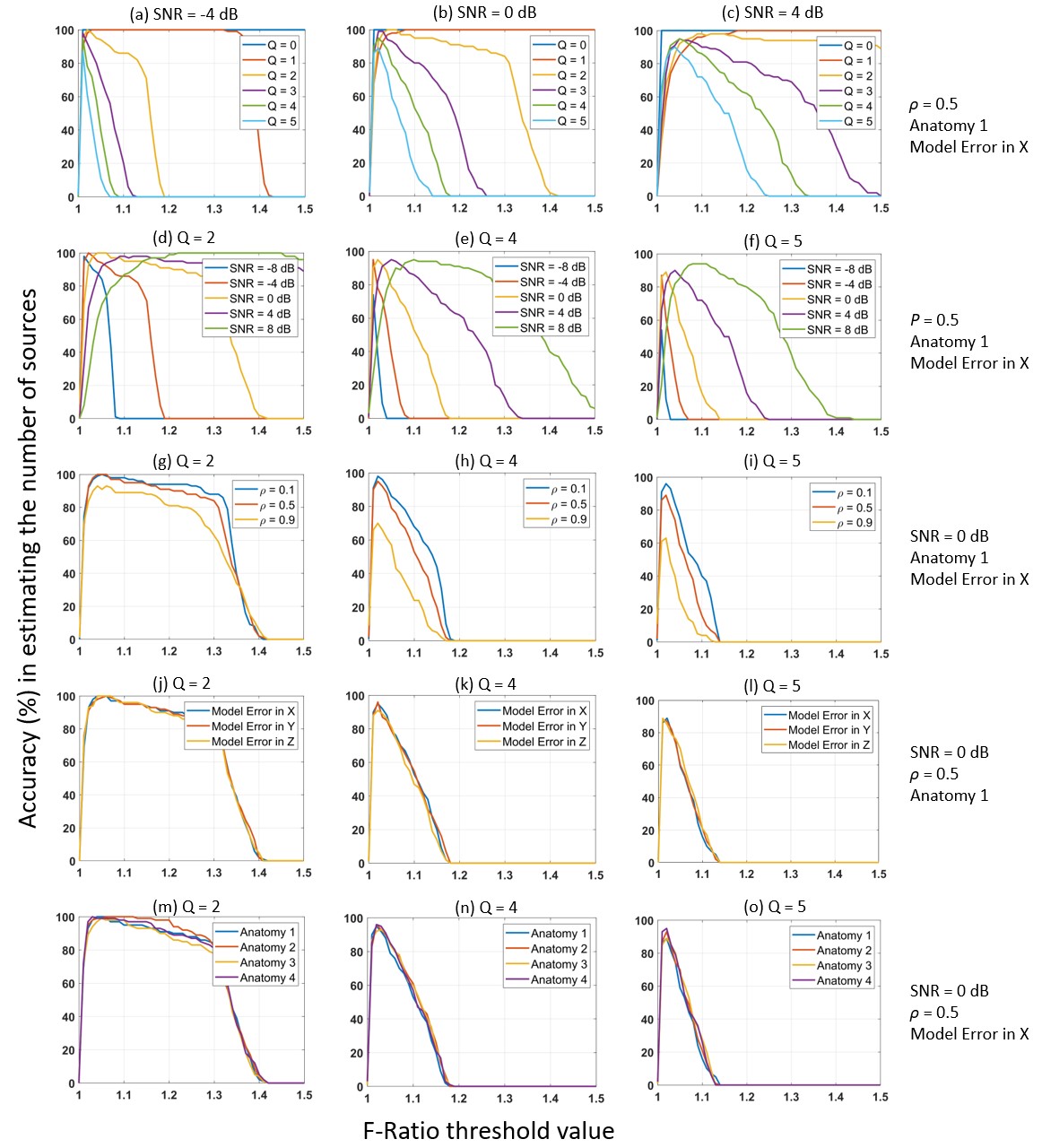}
\end{center}
\caption{Accuracy of the F-ratio method for estimating the number of active MEG sources under different thresholds. Performance evaluation was conducted across various experimental conditions: (a-c) varying number of true sources $Q$, (d-f) SNR levels from -8 to 8 dB, (g-i) inter-source correlation values $\rho$ from 0.1 to 0.9, (j-l) different model errors, and (m-0) different cortical anatomies. Each experimental condition was tested using 100 Monte-Carlo repetitions to ensure statistical robustness. To account for head registration errors, we incorporated inaccuracies into the lead field matrix by applying a translation of 1mm posterior (X-axis), rightward (Y-axis), and upward (Z-axis).} \label{fig:01}
\end{figure}

We observed that the accuracy of estimating the true number of sources using a specific F-ratio threshold was strongly influenced by the actual number of sources $Q$. This relationship is depicted in Figure \ref{fig:01}(a-c), where the estimation accuracy varied significantly for different values of $Q$. Notably, as the number of true sources increased, a lower F-ratio threshold was required to achieve higher performance. Similarly, we discovered a strong correlation between the accuracy of estimating the true number of sources and the SNR level. Figure \ref{fig:01}(d-f) demonstrates this dependency for various SNR values. As the SNR level increased, a lower F-ratio threshold became necessary to achieve higher accuracy in estimating the true number of sources.

In contrast, we made the important observation that the optimal F-ratio threshold remained independent of the inter-source correlation level. This finding is illustrated in Figure \ref{fig:01}(g-i), where we tested different inter-source correlation values (0.1, 0.5, and 0.9). The accuracy in estimating the number of sources peaked at the same threshold value for all correlation levels. This robustness indicates that the optimal F-ratio thresholds were not influenced by the inter-source correlation values of the active sources. Consequently, researchers may rely on a consistent threshold value regardless of the degree of correlation among the sources, enhancing the practical applicability and reliability of the F-ratio method. We obtained similarly robust results when testing the accuracy of estimating the true number of sources across different model errors (Figure \ref{fig:01}(j-l) and cortical anatomies (Figure \ref{fig:01}(m-o). In the case of model errors, we introduced registration errors by translating the reconstruction source space relative to the source simulation space. Specifically, we applied translations of 1 mm posterior (X-axis), right (Y-axis), and upward (Z-axis). Importantly, despite the presence of these registration errors, the F-ratio method remained highly robust. Similarly, when evaluating the F-ratio thresholds across the cortical anatomies of four different adult humans, we observed consistent and robust results. 

In summary, our observations indicated that the performance of F-ratio thresholds varied significantly depending on the true number of sources and the SNR levels of the data. However, we found that F-ratio thresholds remained robust across different inter-source correlation values, model errors, and cortical anatomies. These findings highlight the need to adapt and optimize threshold procedures for the F-ratio test based on the specific number of sources and SNR levels in the data. In the next section, we determined these optimal thresholds.

\begin{figure}[b]
\begin{center}
\includegraphics[width=0.5\linewidth]{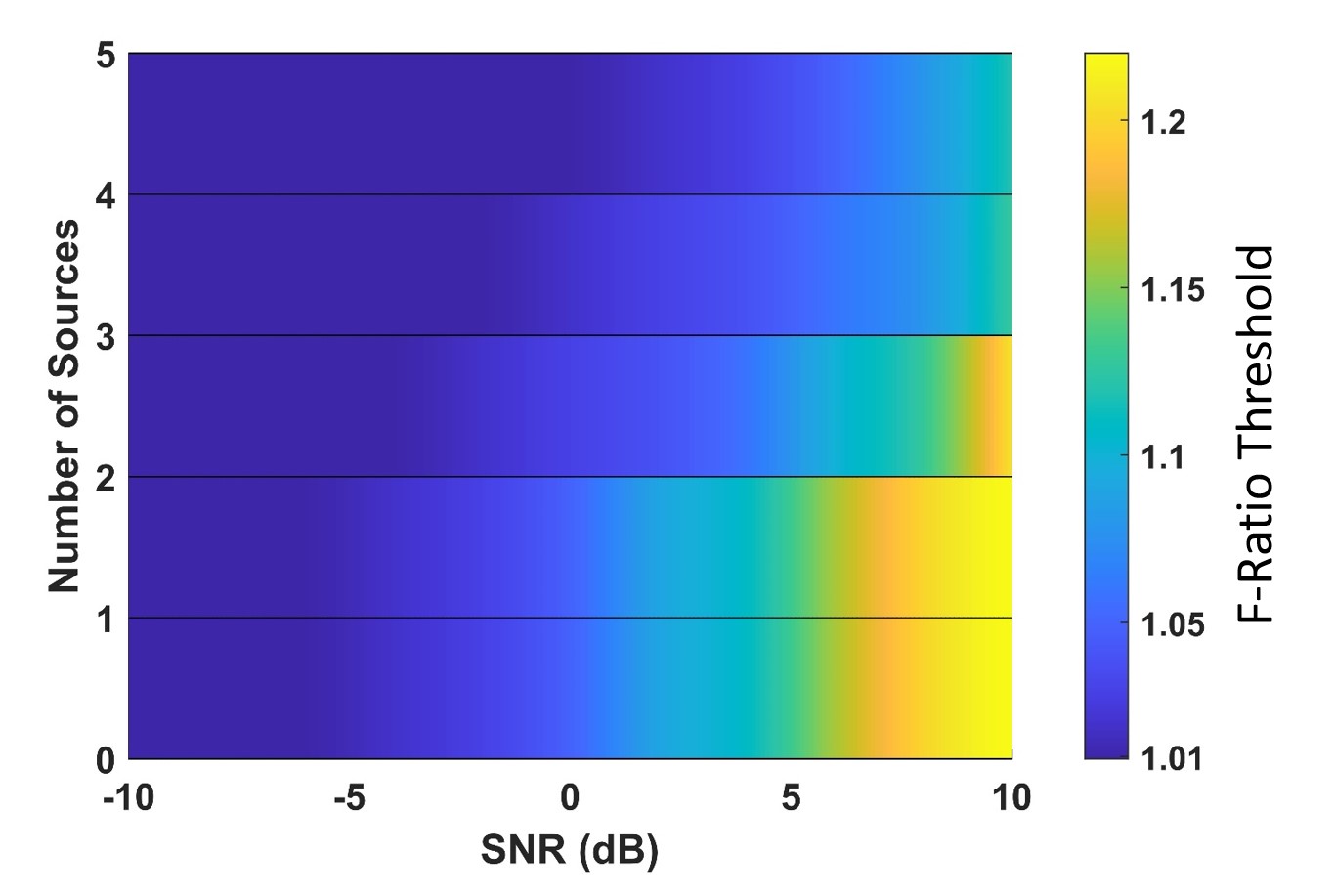}
\end{center}
\caption{Optimal F-ratio threshold values, adjusted for the signal-to-noise ratio (SNR) level and the number of active sources in the MEG data.}\label{fig:02}
\end{figure}
\begin{figure}[b]
\begin{center}
\includegraphics[width=\linewidth]{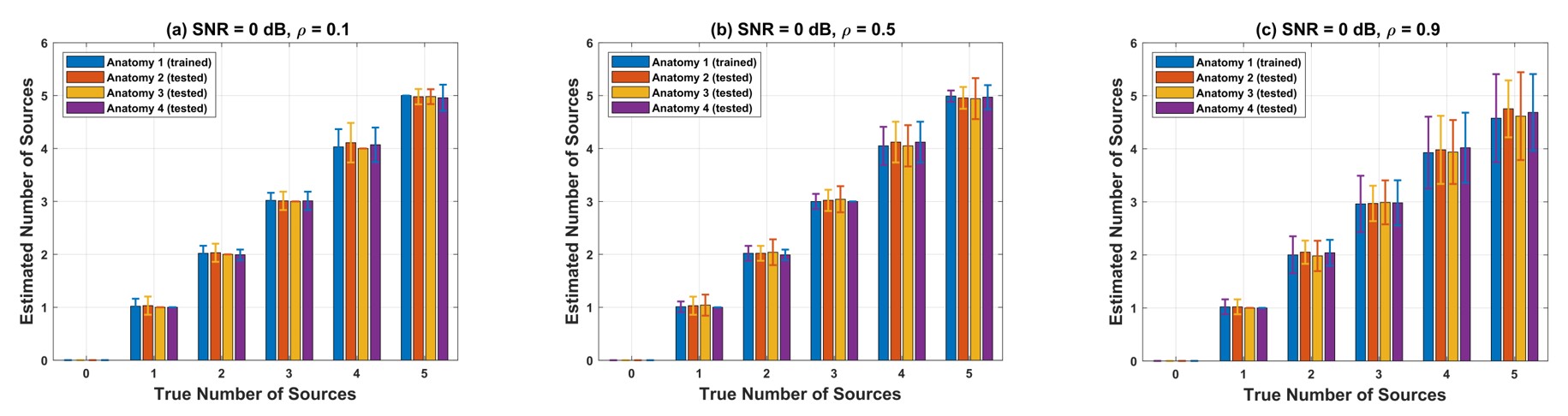}
\end{center}
\caption{Performance of adjusted F-ratio thresholds in simulated data. The thresholds were optimized for Anatomy 1 and applied to three different anatomies (Anatomies 2, 3, and 4). The F-ratio method was employed to estimate the number of sources under varying levels of source correlation (a) $\rho$ = 0.1, (b) $\rho$ = 0.5, and (c) $\rho$ = 0.9, while maintaining a signal-to-noise ratio of 0 dB.}\label{fig:03}
\end{figure}

\subsection{Computation and evaluation of adjusted F-ratio thresholds in simulated data}

In this section, we aimed to determine adjusted F-ratio thresholds for accurately estimating the number of active sources in MEG data. To accomplish this, we used a specific cortical anatomy as a reference (referred to as Anatomy 1) . Optimal threshold values for Anatomy 1 were computed by identifying the F-ratio value that yielded the highest average accuracy in estimating the number of sources, considering the following experimental conditions: 1mm modeling errors in $x, y, z$ and inter-source correlations $\rho \in \{0.1, 0.5, 0.9\}$. The resulting adjusted F-ratio thresholds were obtained for various SNR levels and numbers of sources, as depicted in Figure \ref{fig:02}. Our findings indicate that higher threshold values are necessary in scenarios characterized by a high SNR and a low number of sources.

To assess the effectiveness of the adjusted threshold values obtained for the reference anatomy, we conducted tests on three additional cortical anatomies (Figure \ref{fig:03}). The performance of the adjusted F-ratio thresholds in estimating the number of sources was evaluated at 0 dB SNR and correlation levels $\rho \in \{0.1, 0.5, 0.9\}$. Remarkably, the results showed that the performance of the adjusted thresholds was comparable to that of the reference cortical anatomy (Anatomy 1). These findings demonstrate the reliability and robustness of the calculated optimal F-ratio thresholds across a wide range of simulation scenarios, including variations in the number of sources, SNR levels, inter-source correlation values, modeling errors, and cortical anatomies.

\begin{figure}[t]
\begin{center}
\includegraphics[width=\linewidth]{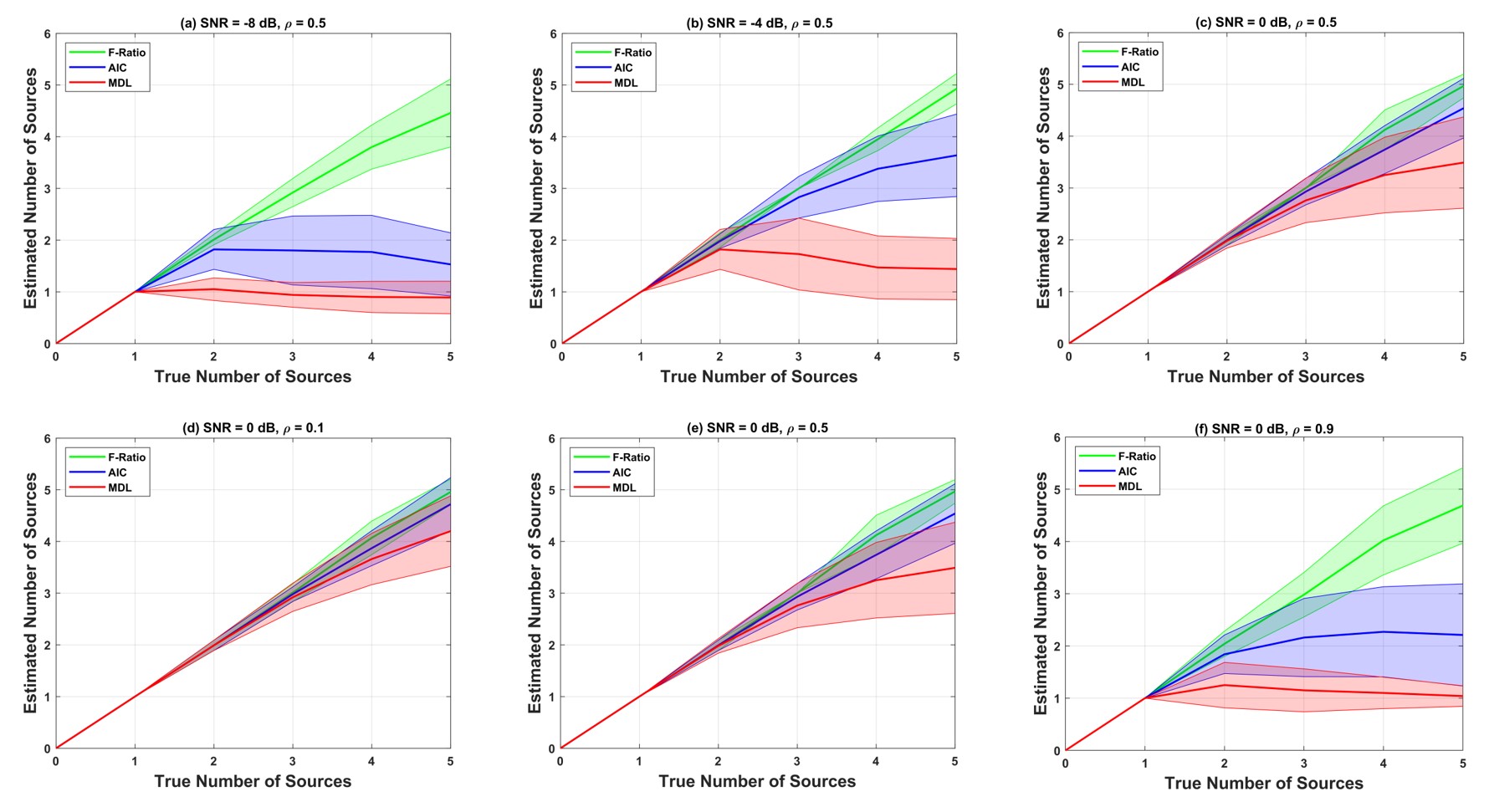}
\end{center}
\caption{Comparison of the F-ratio, AIC, and MDL methods for estimating the true number of dipoles at various SNR levels: (a) SNR = -8 dB, (b) SNR = -4 dB, and (c) SNR = 0 dB, and different levels of source correlation: (d) $\rho$ = 0.1, (e) $\rho$ = 0.5, and (f) $\rho$ = 0.9.}\label{fig:04}
\end{figure}

We proceeded by conducting a comparative analysis between the proposed F-ratio method with adjusted thresholds and two commonly used methods, namely the information criterion AIC and MDL method, for estimating the number of sources. These methods rely on likelihood functions derived from information theory to assess and choose the optimal model. The comparison results of the F-ratio, AIC, and MDL methods across different SNR conditions and correlation levels are depicted in Figure \ref{fig:04}. The results  correspond to cortical anatomy 4 with 1 mm modeling error in X. Remarkably, the proposed F-ratio method with adjusted thresholds outperformed both the AIC and MDL methods in terms of accuracy and reliability. 

\subsection{Performance of the F-ratio method in estimating the number of active dipoles in phantom data}

\begin{figure}[b]
\begin{center}
\includegraphics[width=\linewidth]{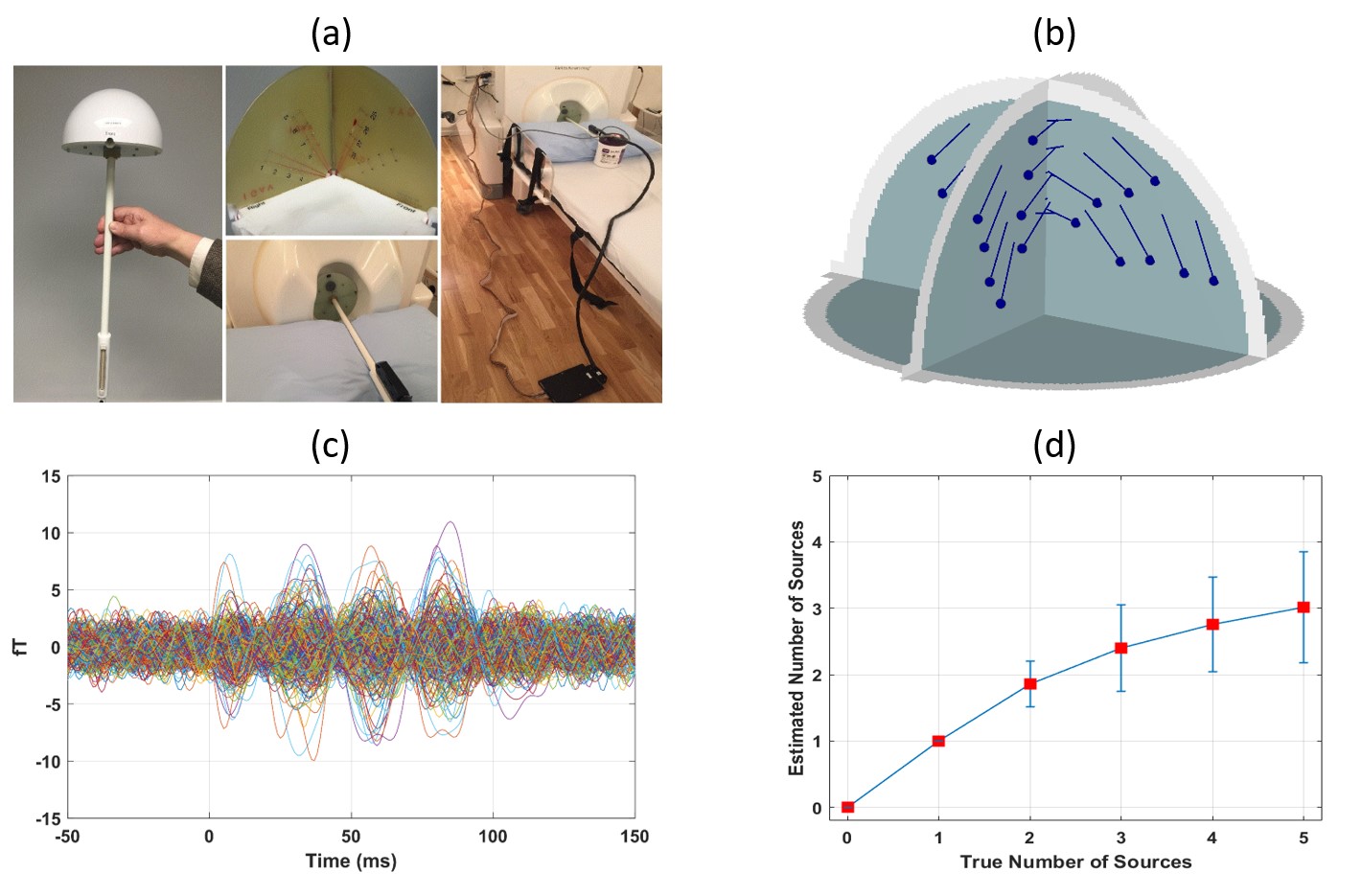}
\end{center}
\caption{Performance of F-ratio method in phantom data. (a) Real phantom provided by the MEG vendor MEGIN \citep{phantomtutorial}. (b) Location of the 32 artificial dipoles of the MEGIN phantom. (c) Example sensor measurements from two active dipoles with a temporal delay of 29 ms, following temporal and spatial prewhitening, corresponding to SNR 5.5 dB. (d) Performance of F-ratio method in estimating the number of active dipoles.}\label{fig:05}
\end{figure}

We assessed the performance of the F-ratio method in estimating the number of active dipoles in phantom data (Figure \ref{fig:05}a). The locations of the 32 artificial dipoles of the MEGIN phantom are shown in Figure \ref{fig:05}b. To simulate the activation of multiple MEG sources simultaneously, we combined the data obtained from individually activated dipoles. To avoid perfect coherence, a random delay ranging from 0 to 50 ms was introduced between the dipole time courses. Figure \ref{fig:05}c illustrates an example of MEG sensor data from two active dipoles with a randomly selected temporal delay of 29 ms. The time courses are displayed following temporal and spatial prewhitening, and had an estimated SNR of 5.5 dB.

We conducted 100 Monte Carlo repetitions of phantom data simulations for each scenario involving 0 to 5 active sources. For each repetition, we applied the adjusted F-ratio thresholds based on the estimated SNR of the data and the corresponding number of tested sources. The performance of the F-ratio method in accurately estimating the true number of active dipoles is shown in Figure~\ref{fig:05}d. The method successfully identified the correct number of sources up to 2, surpassing both the AIC and MDL methods. In contrast, the latter methods failed entirely, with no correct estimations among the 100 simulated scenarios (results not depicted). 

It is important to note that although the performance of the F-ratio method in experimental data was not as remarkable as in the simulated data, we attribute this to two factors. First, the specific configuration of the phantom dipoles played a critical role. The 32 phantom dipoles were closely spaced and shared similar orientations. In the 100 Monte Carlo repetitions, sources were randomly chosen with no constraints in their simultaneous activation. Consequently, there was a substantial probability of selecting adjacent sources, and this likelihood increased as the number of sources grew (3, 4, and 5). Additionally, the time courses of the phantom dipoles had random delays ranging from 0 to 50 ms, resulting in instances of minimal delay and strong correlation. The combined effect of proximate source selection and small time course delay exacerbated the challenge of accurately estimating the number of sources, particularly when dealing with a larger number of sources. Localization errors are known to increase when dealing with sources that have small spatial separation. This phenomenon has also observed in the AP method, as we reported in our previous work \cite{adler2019localization}. Second, estimating the number of sources beyond two is generally an exceptionally challenging problem in MEG. Consequently, a significant body of work has focused on solving the problem of localizing up to two or three sources within simulation or controlled environments \citep{mosher_multiple_1992,Mosher1999,makela_truncated_2018,ilmoniemi2019brain,adler2019localization,giri2020brain}.

\subsection{Performance of the F-ratio method in estimating the number of sources in human auditory data}

To evaluate the effectiveness of the F-ratio method in human data, we utilized it to analyze brain responses captured during an auditory task. We employed the AP source localization method to fit dipoles within the time interval of 100-130ms relative to the onset of the auditory stimuli, which approximately corresponded to the peak MEG response. The SNR within this interval was determined to be 5.78 dB. 

\begin{figure}[t]
\begin{center}
\includegraphics[width=\linewidth]{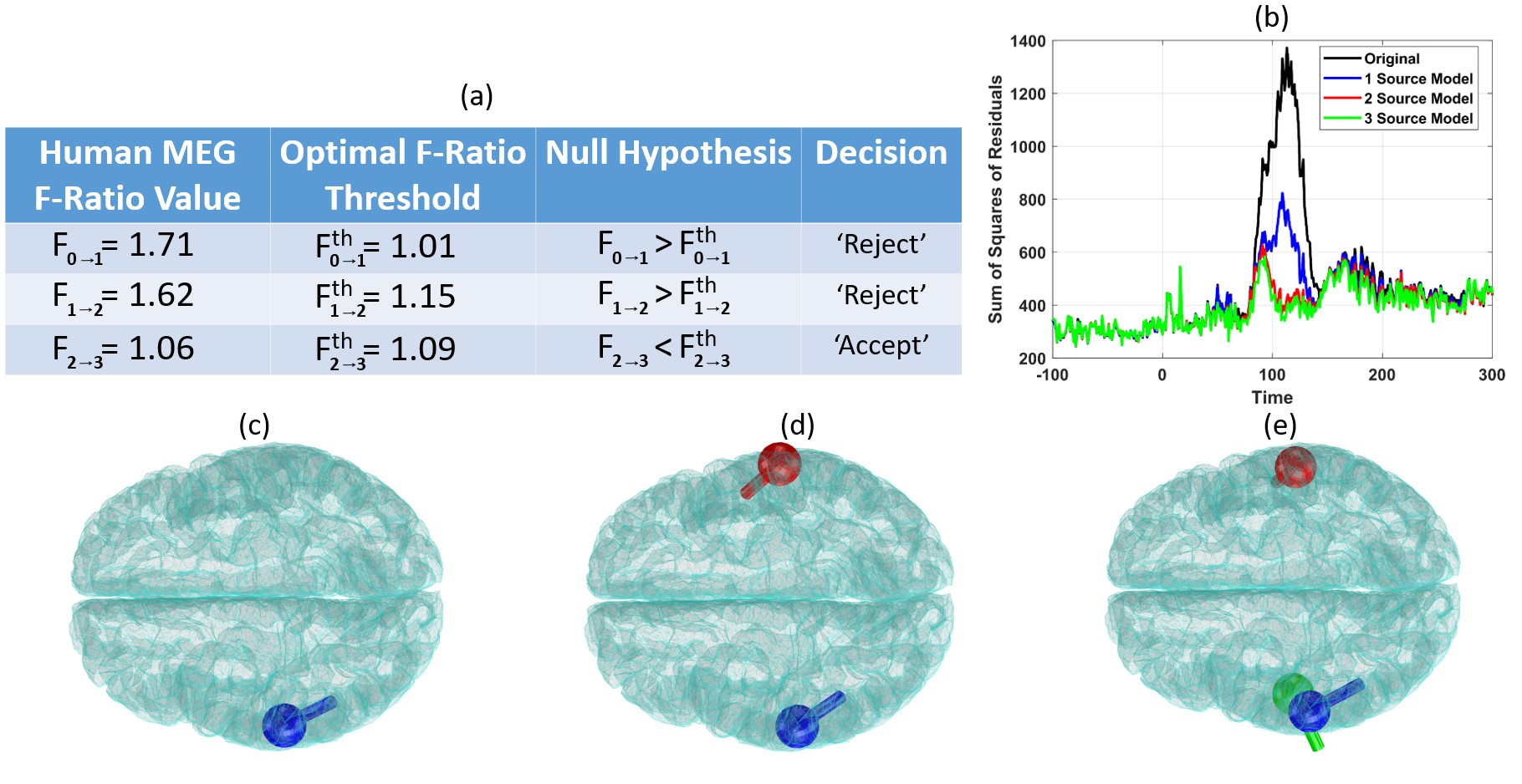}
\end{center}
\caption{Performance of the F-ratio method in estimating the number of sources in human auditory data. (a) Comparison of obtained F-ratio values in human data with optimal F-ratio thresholds, supporting a model with two active sources. (b) Plot of the sum of squares of residuals for models with different numbers of active sources. (c) Localization of a single source. (d) Localization of two sources. (e) Localization of three sources.}\label{fig:06}
\end{figure}

Figure \ref{fig:06}a illustrates the adjusted F-ratio thresholds at 5.78 dB SNR, along with the corresponding values estimated from the human data. By employing the F-ratio sequential procedure, we observed the rejection of the reduced models with zero or one source, while finding no evidence to reject (and thus accepting) a model with two distinct sources. To further validate these findings, we plotted the sum of squares of residuals for the competing models, as shown in Figure \ref{fig:06}b. This plot reinforces our conclusion of the presence of two sources, as there was no significant reduction in the sum of squares of residuals beyond the model order of 2. Additionally, we visualized the dipoles detected using the AP localization method for the cases of 1, 2, and 3 sources (Figure \ref{fig:06}c-e). In the case of the two-source model, the dipoles were localized bilaterally and coincided with the well-established regions in the primary auditory cortices.

It is worth mentioning that alternative methods such as the AIC and MDL yielded different estimates for the number of sources. Specifically, the AIC method suggested 33 sources, while the MDL method indicated only 31 sources. These estimates far exceed the expected number of active sources in human auditory responses and are not in line with the existing knowledge in the field.

\section{Conclusion}

We have validated our F-ratio-based method on simulated, real phantom, and human MEG data. In comparison to other state-of-the-art statistical approaches like AIC and MDL, which rely on certain assumptions that often do not hold in real-world situations, our method demonstrated superior performance in terms of accuracy and reliability.

One crucial aspect we emphasized is the selection of appropriate thresholds for the F-ratio values, which significantly impacts the overall performance of the method. We identified optimal thresholds and showed that these thresholds needed to be adjusted for the number of sources and SNR levels. Notably, these thresholds exhibited remarkable consistency across different inter-source correlations, modeling errors, and cortical anatomies. Overall, by fine-tuning the selection of thresholds, our F-ratio-based method provides researchers with a precise and robust tool for accurately estimating the true number of active sources in MEG data.

However, it is crucial to acknowledge that the adjusted threshold values obtained in this study are specific to the MEG system analyzed and may need to be adjusted for other devices or modalities, such as EEG. When applying the F-ratio method in different devices, it would be necessary to determine appropriate threshold values that are specific to each case. Similarly, the effectiveness of the proposed method is inherently linked to the choice of the source localization technique. In this study, we employed the AP method to compute and validate the F-ratio thresholds. However, different source localization methods may yield varying results and require different threshold adjustments. Therefore, it is crucial to determine the optimal threshold values for a particular set of experimental settings and source localization method to ensure accurate estimation of the true number of active brain sources.

Despite these considerations, our proposed method provides researchers with a precise tool to estimate the true number of active brain sources and effectively model brain function. By calculating threshold values that are tailored to the specific modality and source localization method, researchers can enhance the accuracy and reliability of their source estimation process. Further research is needed to explore and validate the proposed method in different modalities and with various source localization techniques. By refining the threshold determination process and investigating its applicability across different experimental conditions, we can extend the utility of this method to a wider range of neuroimaging studies and enhance our understanding of the underlying mechanisms of brain function.

\section*{Conflict of Interest Statement}
The authors declare that the research was conducted in the absence of any commercial or financial relationships that could be construed as a potential conflict of interest.

\section*{Author Contributions}
All authors contributed to conception of the method. AG and DP developed the methodology. AG performed the coding and data analysis. All authors interpreted and discussed the results. AG wrote the first manuscript draft. DP made major contributions to the writing of the final manuscript. AA and DP provided funding, resources and supervision. All authors critically reviewed the manuscript and approved the submitted version.

\section*{Funding}
This work was supported by NIH grant 1R01EY033638-01 (to DP) and the United States-Israel Binational Science Foundation grant 2020805 (to AA). 

\section*{Publisher's note}
All claims expressed in this article are solely those of the authors and do not necessarily represent those of their affiliated organizations, or those of the publisher, the editors and the reviewers. Any product that may be evaluated in this article, or claim that may be made by its manufacturer, is not guaranteed or endorsed by the publisher.


\section*{Ethics statement}
The study was approved by the local ethics committee (Institutional Review Board of the Massachusetts Institute of Technology), following the principles of the Declaration of Helsinki. The patients/participants provided their written informed consent to participate in this study.


\section*{Data Availability Statement}
Data and code for our implementations will be made available at \href{https://github.com/Amita-Giri/FRatio-based-Method-for-Source-Enumeration} {https://github.com/Amita-Giri/FRatio-based-Method-for-Source-Enumeration}.

\bibliographystyle{Frontiers-Harvard} 
\bibliography{main}

\end{document}